\documentclass[journal]{IEEEtran}
\usepackage[flushleft]{threeparttable}
\usepackage{multirow}
\usepackage{amsmath}
\usepackage{amssymb}
\usepackage{graphicx}
\usepackage{gensymb} % for \degree
\usepackage{comment}
\usepackage{subfig}
\usepackage{graphics}
\usepackage{slashbox}

\usepackage[hyphens]{url}
\usepackage{breakurl}
\usepackage{authblk}
\usepackage{hyperref}

\newtheorem{proposition}{Proposition}
\newtheorem{property}{Property}
\begin{document}

\title{Real-time System Identification Using Deep Learning for Linear Processes with Application to Unmanned Aerial Vehicles}

\author[*1]{\textbf{Abdulla Ayyad}}
\author[*1]{\textbf{Mohamad Chehadeh}}
\author[1]{\textbf{Mohammad I. Awad}}
\author[1,2]{\textbf{Yahya Zweiri}}
\affil[1]{Khalifa University Center for Autonomous Robotic Systems, Khalifa University, Abu Dhabi United Arab Emirates}
\affil[2]{Faculty of Science, Engineering and computing, Kingston University, London SW15 3DW, UK}
\affil[*]{Equal contribution authors}

\maketitle

\begin{abstract}
This paper proposes a novel parametric identification approach for linear systems using Deep Learning (DL) and the Modified Relay Feedback Test (MRFT). The proposed methodology utilizes MRFT to reveal distinguishing frequencies about an unknown process; which are then passed to a trained DL model to identify the underlying process parameters. The presented approach guarantees stability and performance in the identification and control phases respectively, and requires few seconds of observation data to infer the dynamic system parameters. Quadrotor Unmanned Aerial Vehicle (UAV) attitude and altitude dynamics were used in simulation and experimentation to verify the presented methodology. Results show the effectiveness and real-time capabilities of the proposed approach, which outperforms the conventional Prediction Error Method in terms of accuracy, robustness to biases, computational efficiency and data requirements. 
\end{abstract}

\begin{IEEEkeywords}
System Identification, Unmanned Aerial Vehicles, Robot Learning, Sliding Mode Control, Process Control.
\end{IEEEkeywords}

\section*{Multimedia Material}
A supplementary video for the work presented in this paper can be accessed at: \url{https://www.youtube.com/watch?v=dz3WTFU7W7c}

\section{Introduction}

Since the third industrial revolution, system identification has been a key element in the development of autonomous technologies in a wide set of industrial applications. Accurate knowledge of system dynamics enables the design of robust and high-performance systems for prediction, planning and control. Unmanned Aerial Vehicles (UAVs) are an example of an autonomous system that has seen diverse utilization in areas of agriculture, disaster relief, remote sensing, surveillance, etc. \cite{wu2019, shaffer2018, shakhatreh2019, kim2019, kalyaev2017}. UAVs are often deployed in uncontrolled environments, and are hence required to adapt to dynamic conditions in real-time with minimal sacrifice to functionality and performance. 

To meet the aforementioned requirements of autonomy, extensive research have been carried out to develop effective methods of system identification and adaptation. These methods are generally classified as parametric or non-parametric depending on the control requirements and design constraints. Parametric means are data-driven approaches where model parameters of a Process Under Test (PUT) are identified based on observation data. Such approaches include prediction error methods (PEM) \cite{Ljung1998, Ljung2002, van2013identification}, maximum likelihood (ML) methods \cite{hagenblad2008maximum, tong1998multichannel}, least square (LS) methods \cite{hu2020}, frequency response identification methods \cite{Pintelon1994,  Nevaranta2016, Villwock2008}, and neural network based methods \cite{chen1990non, lu1998robust, delgado1995dynamic}. Several studies in the literature applied these techniques to UAV operation with accurate identification results \cite{Grymin2015, Ahsan2017, Wei2017, Bhandari2017, Patil2015, Jin2016}. Nonetheless, these methods require extensive data generation and accurate selection of optimizer initial conditions, which demand human experience and cause susceptibility to data biases and overfitting. Furthermore, most of these methods are computationally expensive and not suitable for real-time application.

\begin{figure*}[t]
\includegraphics[width=\textwidth]{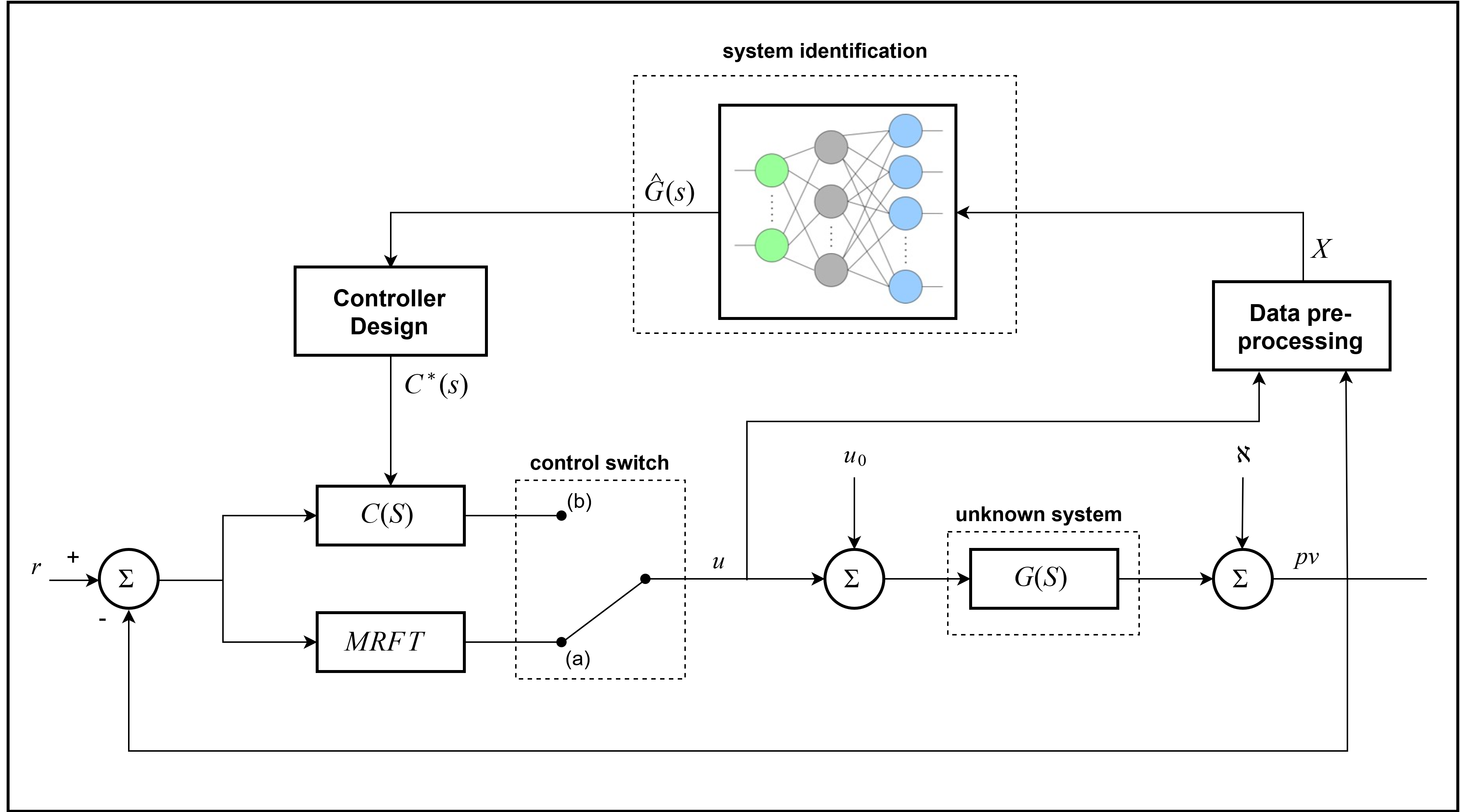} 
\caption{The proposed system identification scheme. The control switch starts at position (a) where MRFT is  used to excite an unknown plant \(G(s)\) with stable oscillations. These oscillations are pre-processed and forwarded to a DL classifier to identify the system parameters as \(\hat{G}(s)\). After the PUT was identified, a suitable controller \(C^{*}(s)\) can be designed and applied to the plant by shifting the control switch to position (b).}
\label{fig:sys_id_overview}
\end{figure*}

On the other hand, non-parametric identification includes methods which rely on the partial knowledge of the PUT to tune a predefined controller structure. Such methods include, for example, the classical Ziegler-Nichols method \cite{ziegler1942optimum}, the Relay Feedback Test (RFT) \cite{Astrom1984}, and the Modified Relay Feedback Test (MRFT) \cite{Boiko2013}. In all these methods, the knowledge of the PUT response to a single excited frequency is enough to get controller parameters tuning. It was shown in recent work \cite{Chehadeh2019} that a near optimal controller for quadcopter attitude dynamics can be designed using MRFT based tuning rules. However, non-parametric methods in the literature are limited to PID tuning and do not provide full insight to the system dynamics as they cannot be used to obtain model parameters. 

Recent advancements in the fields of Iterative Learning Control (ILC), Reinforcement Learning (RL) and Deep Learning (DL), and the growth of computational capabilities have given rise to new approaches of controller design and tuning \cite{Schoellig2012, lambert2019low, Berkenkamp2016, koch2019reinforcement, molchanov2019sim, hwangbo2017control, bansal2016learning}. These approaches have introduced advantages in regards to accuracy of models and controllers, adaptation time, and the ability to handle nonlinearities in the PUT; with the limiting requirement of abundant observation data. Similar to non-parametric tuning, these approaches do not generate explicit estimates of model parameters, but rather implicitly consider these parameters in controller design.

This paper bridges the gap between parametric and non-parametric methods and presents a novel methodology to infer accurate estimates of model parameters with the prime motivation of designing high-performance controllers online and in real-time. The novelty of the proposed methodology lies in utilizing self-excited oscillations (i.e. chattering) resulting from a sliding mode controller, which is the MRFT, to reveal distinguishing information about the PUT. The information revealed by MRFT are then fed to a DL classifier which selects model parameters that best represents the PUT. Fig. \ref{fig:sys_id_overview} illustrates the proposed comprehensive system identification approach. We show that this identification method can be performed in real-time such that a UAV adapts to changes to its own physical dynamics during a flight mission. The suggested online identification methodology is safe with guaranteed stability, and results in controllers with assessable levels of robustness and performance. The presented approach is mainly applied to Second Order with Integrator Plus Time Delay (SOIPTD) linear systems; but is also applicable to other system models with equal or lower number of model parameters. Due to its real-time capabilities, the proposed technique can handle static nonlinearities by applying the identification process in multiple operation modes (e.g. near ground hovering or drag dynamics caused by large translational speeds); to obtain locally linear descriptions of the system.

\renewcommand{\arraystretch}{2}
\begin{table*}[t]
    \caption{Qualitative comparison with selected recent methods from literature addressing the problem of automatic controller tuning and adaptation for UAVs}
    \centering
    \begin{tabular}{|p{3cm}|p{3cm}|p{2cm}|p{3cm}|p{3cm}|}
        \hline
            \textbf{Method} & \textbf{Experimental data generation} &
            \textbf{Computational Resources}  &
            \textbf{Stability} & \textbf{Comments} \\
        \hline
        \hline
        This work & Single steady-state oscillation at a specified phase  &  A few milliseconds with modern on-board processors & Guaranteed in data generation phase by Loeb criterion \cite{loeb1965recent,Boiko2013} & Not investigated for lateral UAV motion yet. Provides PUT model parameters \\
        \hline
        Sim-to-(Multi)-Real using proximal policy optimization (PPO) RL \cite{molchanov2019sim} & Not required  & Inference model is running in real-time & No stability guarantees & This work's problem statement is closest to the one presented in this paper, but with no guarantee of stability or explicit inference of model parameters  \\
        \hline
        ILC \cite{Schoellig2012} & Requires a lot of iterations & Computationally expensive & Stability guaranteed within ILC iterations & Feedforward compensation terms\\
        \hline
        Deep Model-Based Reinforcement Learning \cite{lambert2019low}  & Requires a lot of experimental data  & Computationally expensive & Not guaranteed & Early adaptation of model-based RL\\
        \hline
        Learning through Gaussian processes with Bayesian optimization \cite{Berkenkamp2016}  & Requires a lot of experimental data & Computationally expensive & Guaranteed stability during learning within model uncertainty margins & \\
        \hline
        Non-parametric tuning of UAV inner loops \cite{Chehadeh2019,poksawat2016automatic}& Single steady-state oscillation at a specified phase  & Negligible & Guaranteed stability during identification & Does not provide model parameters\\
        \hline
        Heuristics based tuning \cite{giernacki2019real} & Requires a lot of experimentation data & Low computational resources & Not guaranteed and subject to selected optimization parameters constraints & \\
        \hline
    \end{tabular}
    \label{tab:method_vs_others}
\end{table*}

There are two inherent features of the proposed approach  which highlight its advantages over other existing methods. First, the required amount of data needed for identification is minimal as it consists of a single excited frequency of the system. In contrast, other data driven classical identification methods used in the literature require extensive data generation \cite{Ljung1998,Ljung2002,van2013identification,hagenblad2008maximum,tong1998multichannel,Nevaranta2016,Villwock2008}. Due to the reduction in data requirements, the proposed methodology does not require human experience for data generation, and the model fitting process is not prone to data bias. This makes the proposed method precise and accurate in identifying unknown model parameters. Other advantages of data reduction include minimizing computational requirements and shortening the period of the identification phase. In fact, we have found that the required computational time on modern commercial processors is in the order of milliseconds; and the identification phase lasts for a maximum of a few seconds (this depends on the PUT dynamics, e.g. mass).

The second inherent feature of the presented identification method is the guarantee of stability during the identification phase. This relieves the need for hand-tuned initial stabilizing controllers as opposed to classical identification methods, and makes the presented method less prone to estimation biases and non-optimality caused by the selection of initial control parameters (i.e. initial parameters for the optimizer decision variables). As stability is guaranteed, the PUT can be directly started in the identification phase as demonstrated in the results section; thus further minimizing the required operation time to estimate  model parameter. Table \ref{tab:method_vs_others} provides a qualitative comparison between the method suggested in this paper and other related work in literature. Additionally, the Results section provides a quantitative comparison between this paper's method and two other identification methods: PEM and non-parametric tuning based on MRFT.

The main contributions of this paper can be summarized as:

\begin{itemize}
    \item We introduce a novel approach of parametric system identification with guaranteed stability, real time capabilities, and minimal requirements of observation data.
    \item We optimize the identification phase to reveal distinctive information about the PUT by means of finding the distinguishing phase for a set dynamic systems. 
    \item We present a discretization technique to address system identification as a classification problem by utilizing the concept of controller performance deterioration.
    \item We devise a modified formulation of the Softmax activation function that adds a meaningful discrepancy to the cost of misclassification, leading to faster and more accurate training of the DL model.
\end{itemize}

The remainder of this paper is organized as follows. The system identification problem is first formulated in Section \ref{sec:problem_statement}. Section \ref{sec:methodology} describes the comprehensive identification approach proposed in this paper. In section \ref{sec:results}, simulation and experimental results for the suggested approach are presented, discussed and compared against PEM based system identification and the non-parametric tuning rules of \cite{Chehadeh2019}. Finally, Section \ref{sec:conclusion} summarizes the findings of this paper and provides concluding remarks.

\section{Problem Statement} 
\label{sec:problem_statement}

Considering an LTI system \(G(s)\) with known model structure and unknown set of bounded model parameters \(\vec{p}\in (D \subset \mathbb{R}^Q)\) where \(Q\) corresponds to the number of unknown model parameters in \(G(s)\). Let us also assume a feedback controller \(C(s)\) that acts on \(G(s)\). Given vector \(X \in \mathbb{R}^{2N}\) that contains \(N\) uniformly sampled measurements of both the process variable \(pv\) and controller output \(u\) signals. Inner system states are considered to be unobservable. We wish to find the mapping \(\Gamma: X \rightarrow \bar{D}\) where \(\bar{D}\) is a discretization of the subspace \(D\). \\
In this work, we limit the order of the LTI system to SOIPTD, which corresponds to multirotor attitude and altitude dynamics as presented in \cite{Chehadeh2019}:
\begin{equation}\label{eq_attitude_model_topdt}
G(s)=\frac{K_{eq}e^{-\tau s}}{s(T_{prop}s+1)(T_{body}s+1)}
\end{equation}
where \(K_{eq}\) is the overall lumped gain of the system, \(\tau\) is the overall observed delay in the system, \(T_{prop}\) is the time constant associated with propulsion dynamics, and \(T_{body}\) is the time constant associated with body dynamics. These dynamics relate motor commands sent by the flight controller to the observed roll, pitch, or altitude. The considered attitude and altitude dynamics are subject to measurement noise \(\aleph\) and forced bias \(u_0\) caused by external disturbances (e.g. gravity) or sensors bias. 

\section{Methodology} \label{sec:methodology}

\subsection{Finding the Distinguishing Phase} \label{findphase}

The identification method presented in this paper builds on two propositions.
\begin{proposition}
There is a distinguishing phase \(\varphi_{d}\) at which the sustained self-excited oscillation characteristics can be used to identify the corresponding processes in \(\bar{D}\).
\end{proposition}
\begin{proposition}
The distinguishing phase \(\varphi_{d}\) corresponds to the optimal tuning rules of \cite{Boiko2013}. As such, \(\varphi_{d}\) can be determined by the process of designing optimal non-parametric tuning rules as outlined in \cite{Boiko2013,Boiko2014}.
\end{proposition}

For completeness, we summarize the steps to obtain optimal non-parametric tuning rules as follows \cite{Chehadeh2019,Boiko2013,Boiko2014}:
 \begin{enumerate}
     \item Select process model and the range of the normalized model parameters.
     \item Discretize the selected range of model parameters to a finite set of dynamics processes.
     \item Select tuning rule specifications based on gain margin or phase margin requirements.
     \item Generate a locally optimal tuning rule for every process in the range.
     \item Apply every locally optimal tuning rule to all other processes in the range. Note the performance deterioration for every process due to the application of the non-optimal tuning.
     \item Select the tuning rule with the least deterioration in performance as the global optimum.
 \end{enumerate}

In our case, the model structure is SOIPTD; therefore, the vector \(\vec{p}\) that contains the model parameters is defined as:
 \begin{equation}\label{eq_decison_vec}
\vec{p}=
\begin{bmatrix}
T_{prop}\;
T_{body}\;
\tau
\end{bmatrix}^T
,\quad \vec{p} \in \!R^3
\end{equation}
 and the model parameters range is selected to be 
 \begin{equation}\label{eq_parameter_range_vals}
\begin{bmatrix}
0.015,\,
0.2,\,
0.0005
\end{bmatrix}^T \leq \vec{p} \leq \begin{bmatrix}
0.3,\,
2,\,
0.1
\end{bmatrix}^T\\
\end{equation}
which includes a wide variety of multirotor UAV designs and sizes from racing quadrotors to larger multirotor UAVs having a takeoff weight of up to 50Kg.

Using the aforementioned method to generate optimal non-parametric tuning rules, we found $\varphi_{d}$ to be $-46.89\degree$. In practice, a self-sustained oscillation with a specific phase can be excited using MRFT. MRFT is an algorithm that can be realized with the following equation \cite{Boiko2013}:
\begin{multline}\label{eq_mrft_algorithm}
u_M(t)=\\
\left\{
\begin{array}[r]{l l}
h\; &\text{if}\; e(t) \geq b_1\; \text{or}\; (e(t) > -b_2 \;\text{and}\; u_M(t-) = \;\;\, h)\\
-h\; &\text{if}\; e(t) \leq -b_2 \;\text{or}\; (e(t) < b_1 \;\text{and}\; u_M(t-) = -h)
\end{array}
\right.
\end{multline}
where \(b_1=-\beta e_{min}\) and \(b_2=\beta e_{max}\). \(e_{max}>0\) and \(e_{min}<0\) are respectively the last maximum and minimum values of the error signal after crossing the zero level; and \(u_M(t-)=lim_{\epsilon\rightarrow0^+ }u_M(t-\epsilon)\) is the previous control signal. Prior to the start of MRFT the maximum and minimum error values are set as: \(e_{max}=e_{min}=0\). \(\beta\) is a constant parameter that dictates the phase of the excited oscillations as:
\begin{equation}
    \varphi = \arcsin{(\beta)}
\end{equation}

Using the Describing Function (DF) method, it could be shown that the MRFT achieves oscillations at a specified phase angle by satisfying the Harmonic Balance (HB) equation \cite{atherton1977}:
\begin{equation}\label{eq_hb}
N_d(a_0)W_p(j\Omega_0)=-1
\end{equation}
where \(N_d\) is the DF, \(W_p\) is the process under test, \(a_0\) and \(\Omega_0\) are the amplitude and frequency of the steady state oscillations, respectively. The DF of MRFT is presented in \cite{Boiko2013} as:
\begin{equation}\label{eq_mrft_df}
N_d(a_0)=\frac{4h}{\pi a_0}(\sqrt{1-\beta^{2}}-j\beta)
\end{equation}

The DF method provides an approximate solution that is valid only if \(W_p(s)\) has sufficient low pass filtering properties. It is worth mentioning that the MRFT control signal \(u_M(t)\) has a phase lead relative to the error signal \(e(t)\) in the case of \(\beta<0\), and lags in the case of \(\beta>0\). The MRFT DF intersects the Nyquist plot in the second quadrant for \(\beta<0\); while this intersection occurs in the third quadrant when \(\beta>0\). The Relay Feedback Test (RFT) \cite{Astrom1984} could be thought of as a special case of the MRFT algorithm where \(\beta=0\). For our case, the value of the MRFT parameter \(\beta\) that corresponds to \(\varphi_{d}=-46.89\degree\) is \(\beta_{d}=sin(\varphi_{d})=-0.73\).

\subsection{Discretization of System Parameters' Subspace} 
\label{discretization}

System identification has been generally considered in the literature as a regression problem \cite{Kopp1963, Jansson1996, Juang2010}. However, training a deep learning regression model can raise several instability and complexity concerns as suggested in \cite{Rothe2016}, where a DL network was used for age prediction. To avoid these shortcomings with DL regression, the system identification conundrum in this study is formulated as a classification problem by discretizing the parameter space \(D\) into \(N\) unique sets of system parameters \(\bar{D}=\{G_{1}, G_{2}, ..., G_{N} \}\). System identification hence becomes the problem of selecting a candidate set of process parameters \(G_{i}\) within \(\bar{D}\) that best resembles the dynamics of the ground truth dynamic system \(G_{act}\).

A trivial approach to obtain \(\bar{D}\) would be discretizing \(D\) based on an equispaced distance of the model parameters \(\vec{p}\). Assuming that the equi-space distance was small enough to represent all the pivotal processes, \(\bar{D}\) would end up being an over-discretized representation of \(D\). Adjacent processes of a given subspace of \(\bar{D}\) would have similar frequency response characteristics while adjacent processes in another subspace of \(\bar{D}\) would have vastly different frequency response characteristics. Thus, a trained classifier would be biased towards the subspace where adjacent processes have similar frequency response characteristics. Therefore,  a  meaningful  criterion for discretization must be developed to ensure a balance between  the distinguishability of these processes (i.e. in terms of their frequency response characteristics) and their accuracy in representing \(D\). For this purpose, a joint cost function is introduced based on the concept of controller performance deterioration. To illustrate such joint cost function, let us consider \(\{G_i(s),G_j(s)\}\in \bar{D}\); the joint cost associated with applying \(C_i(s)\), which is the optimal controller of process \(G_i(s)\), to the process \(G_j(s)\) would be given by:
\begin{equation}\label{eq_performance_deterioration}
J_{ij} = \frac{J(C_i(s), G_j(s)) - J(C_j(s), G_j(s))}{J(C_j(s), G_j(s))} \times 100 \%
\end{equation}
where \(J\) is a cost function relating a controller \(C(s)\) to a process \(G(s)\) (i.e. IAE, ISE, etc.). The self-joint cost is defined by the case of \(i=j\), where \(J_{ij}=0\). For the case where \(i\neq j\), \(J_{ij} > 0\) by definition. It must be noted that the joint cost function is non-commutative, i.e. \(J_{ij}\neq J_{ji}\). Therefore, \(J_{(ij)}=max\{J_{ij}, J_{ji}\}\) is used as the discretization criteria to provide a performance guarantee among adjacent members of \(\bar{D}\). In this paper, we have chosen ISE as a system performance index. ISE cost function is given by:
\begin{equation}\label{eq_ise}
J_{ISE}(C, G) = \frac{1}{T_s}\int_{0}^{T_s} e(C,G)^2 dt
\end{equation}
where in this paper we use \(J:=J_{ISE}\) for convenience.

\begin{figure}
\includegraphics[width=\linewidth, trim=0.1 0.1 0.1 0.1]{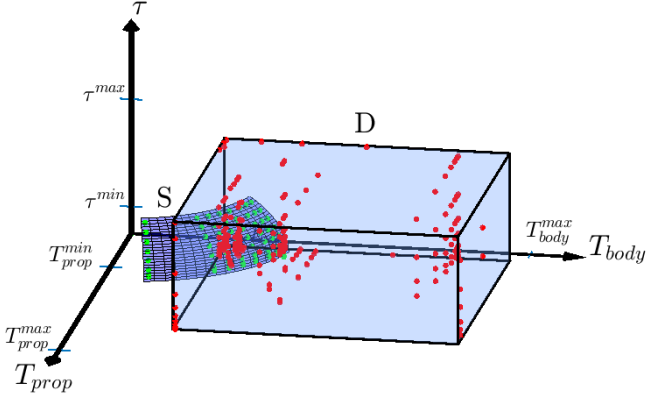} 
\caption{Showing a representation of \(D\) and \(S\) with discretized processes in \(\bar{S}\) shown in green and discretized processes in \(\bar{D}\) shown in red. The discretized processes in \(\bar{D}\) are denser at parts of \(D\) where the ratio of \(\frac{\tau}{T_{prop}}\) and \(\frac{\tau}{T_{body}}\) are highest.}
\label{fig:param_discretization_3d}
\end{figure}

For discretizing \(D\), we choose a desired value of the joint cost between adjacent processes \(J^*\). In this paper, we use \(J^*=10\%\) as it provides sufficient accuracy without requiring excessive simulation time (simulation time have a cubical relationship with the reciprocal of \(J^*\)). To find the process adjacent to a known one, \(G_i\), we use an optimizer (Nelder-Mead simplex algorithm realized by "fminsearch" function in MATLAB$^{\textcircled{R}}$ have been used) that takes a vector of model parameters \(\vec{p}_j\) as the set of decision variables and uses \(E=(J^*-J_{(ji)})^2\) as a cost function. We have found that discretizing \(D\) requires excessive simulation time (might take days to several weeks depending on the selected parameters range of \(D\)). For that we propose reducing the dimensions of the parameters space by transforming the describing subspace \(D\) from rectangular to spherical coordinates. The transformation is given by:
\begin{equation}
    \label{eq_rect_to_sph}
    \begin{array}{c c}
    r_0=\sqrt{T_{prop}^2+T_{body}^2+\tau^2}\\
    \theta=\arctan{(\frac{T_{body}}{T_{prop}})}\\
    \phi=\arccos{(\frac{\tau}{r})}
    \end{array}
\end{equation}

It is worth noting that the parameter \(r\) in \eqref{eq_rect_to_sph} represents time scaling of process parameters \(\vec{p}\) as in \(s'=rs\). This allows us to introduce two properties of the spherical representation that will make the discretization process more efficient.

\begin{property} \label{discretization_property1}
For subsequent time scaling of a system \(G(s)_{i}\) along the radial direction \(G(\alpha s)_{j},\,G(\alpha^{2}s)_{k}\), the joint cost between successive scaled systems remain constant as in: \(J_{ij} = J_{jk},\,J_{ji}=J_{kj}\) for \(\alpha \in \mathbb{R}_{>0}\).
\end{property}

\begin{property} \label{discretization_property2}
Considering two radially scaled systems: \(G(s)_{i}\) and \(G(\alpha_1 s)_{j}\) with a joint cost \(J_{ij}\), and another pair of radially scaled systems \(G(s)_{k}\) and \(G(\alpha_2 s)_{l}\) with the same joint cost \(J_{kl}=J_{ij}\); then \(J^*=J_{(ik)}=J_{(jl)}\) for \({\alpha_1, \alpha_2} \in \mathbb{R}_{>0}\).

%this cost remains constant for any mutual scaling of the two systems along the redial direction \(r\) as follows:
%\begin{equation} \label{discretization_eq1}
%_{G_{(r_{1}, \theta_{1}, \phi_{1})}}J_{G_{(r_{2}, \theta_{2}, \phi_{2})}} = _{G_{(\alpha r_{1}, \theta_{1}, \phi_{1})}}J_{G_{(\alpha r_{2}, \theta_{2}, \phi_{2})}}
%\end{equation}
%For \(\alpha \in R\).
\end{property}

\begin{figure}
\includegraphics[width=\linewidth]{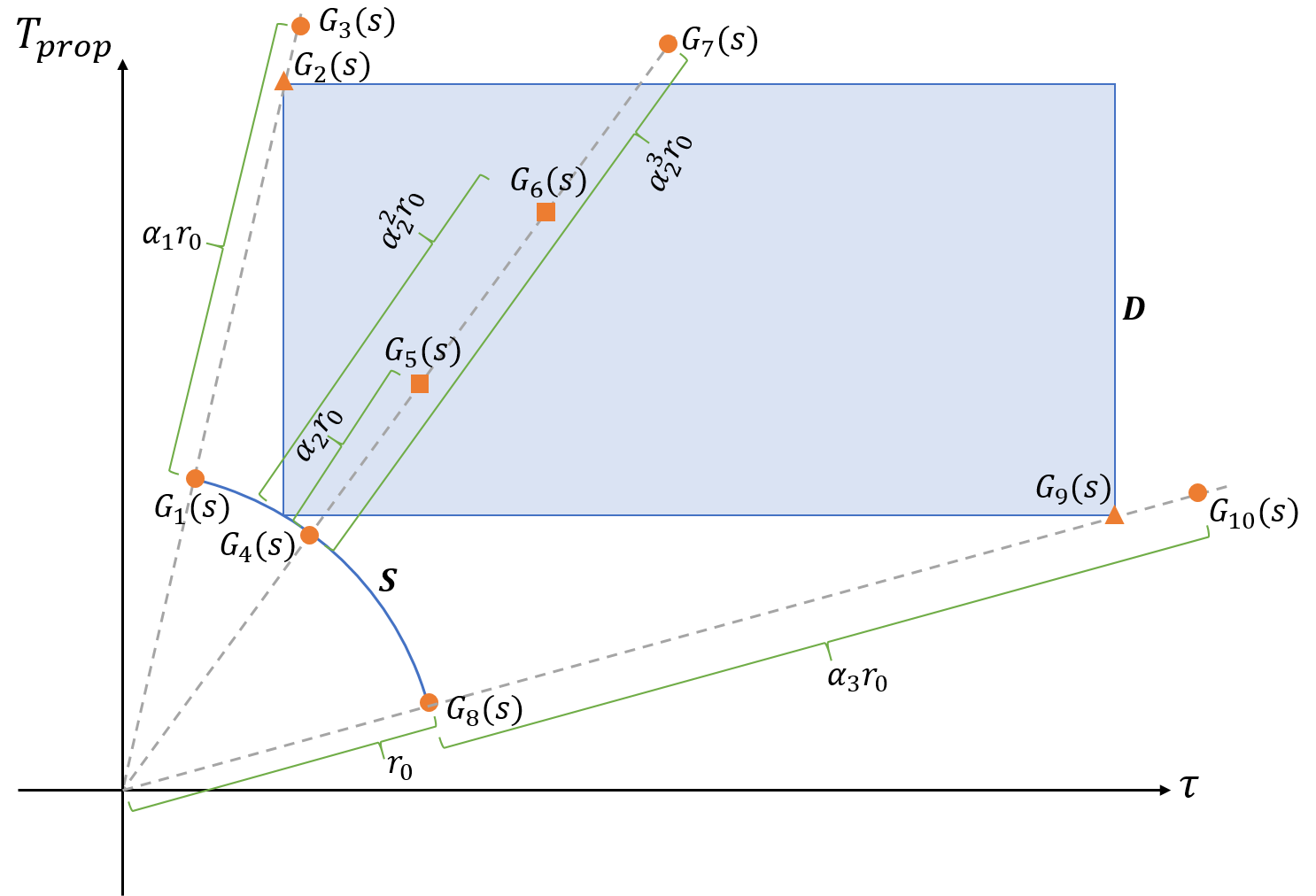} 
\caption{A projected side view showing parameters space \(D\) and the surface \(S\) with radius \(r_0\). \(J^*\) is achieved by using the time scales \(\alpha_1,\alpha_2,\alpha_3\). Note that \(\{G_1(s),G_4(s),G_8(s)\} \in \bar{S}\) and \(\{G_2(s),G_5(s),G_6(s),G_9(s)\} \in \bar{D}\). \(J^{*}=J_{45}=J_{56}\) illustrates property \ref{discretization_property1}. Property \ref{discretization_property2} is illustrated by \(J^{*}=J_{14}=J_{35}\). \(G_2(s)\) and \(G_9(s)\) represent a scaled version of processes \(G_1(s)\) and \(G_8(s)\) respectively  in \(\bar{D}\).}
\label{fig:param_discretization_2d}
\end{figure}

Property \ref{discretization_property1} allows us to discretize a subsurface of a sphere \(S\) that we choose its radius to satisfy \(r_0=||\vec{p}_{min}||\), where \(\vec{p}_{min}\) corresponds to the minimum model parameters set in \(D\). This effectively reduces the discretization problem by one dimension. Fig. \ref{fig:param_discretization_3d} provides a three-dimensional illustration of \(D\) and \(S\). To discretize \(S\) we set \(J^*=10\%\) and we proceed with the discretization by varying the values of \(\phi\) and \(\theta\). To prevent excessive discretization and to increase robustness of controllers against varying model parameters and linearization assumptions, we impose phase margin constraints on controllers used to find joint cost function \(J_{(ij)}\). The phase margin constraint can be imposed using a set of three equations. The first equation relates PID parameters with the PUT amplitude and frequency responses when a steady state oscillation is excited at a certain phase \cite{Boiko2014}:
\begin{equation}\label{eq_homo_tuning_rules}
K_c=c_1\frac{4h}{\pi a_0},\; T_i=c2\frac{2\pi}{\Omega_0},\;T_d=c_3\frac{2\pi}{\Omega_0}
\end{equation}
where \(c_1,c_2\) and \(c_3\) are called the homogeneous tuning rules parameters. These parameters are related with the excitation phase \(\psi_d\) characterized by the MRFT parameter \(\beta\) through the two following equations \cite{Boiko2014}:
\begin{equation}\label{eq_phase_margin_1}
\beta=sin(\phi_m+arctan(\frac{1}{2\pi c_2}-2\pi c_3))
\end{equation}
and:
\begin{equation}\label{eq_phase_margin_2}
c_1\sqrt{1+(2\pi c_3-\frac{1}{2\pi c_2})^2}=1
\end{equation}
where \(\phi_m\) is the imposed phase margin constraint. In this work, we choose \(\phi_m=20\degree\). A higher imposed value of the phase margin constraint will result in lower number of discretized processes. A modified version of Nelder-Mead simplex algorithm that accepts constraints on optimization decision variables is used to realize \eqref{eq_phase_margin_1} and \eqref{eq_phase_margin_2}. Then we proceed by finding \(\bar{S}\) which is the set of the discretized processes in \(S\). Once we have \(\bar{S}\), we find the scaling parameter \(\alpha\) for every process in \(\bar{S}\) as proposed in property \ref{discretization_property1}. Fig \ref{fig:param_discretization_2d} illustrates these steps with the properties \ref{discretization_property1} and \ref{discretization_property2}. Property \ref{discretization_property2} guarantees that all adjacent systems have a joint cost within \(J^*\). Fig \ref{fig:param_discretization_3d} shows the set of discretized processes.

%Once we have found \(\bar{D}\), we can proceed with data generation.
For the parameters range presented in \eqref{eq_parameter_range_vals}, we found the size of \(\bar{D}\) to be 208 processes. The discretized processes are denser at the parts of \(D\) were the ratio between the time delay \(\tau\) and the other process time constants is the highest. Therefore, the identification and control of small UAVs with sensors and actuators that have high delays is found to be more challenging.

\subsection{Deep Learning Model Development and Training} \label{subsec:DNN_model}

In this section, the process of developing and training a deep learning model for system parameter identification is discussed. The objective of the DL model in this study is to find the mapping \(\Gamma: X \rightarrow \bar{D}\) as illustrated in Section \ref{sec:problem_statement}. The input \(X\) to the DL classifier is a uniformly time sampled vector concatenating the controller and plant response while, the output of the DL netwrok is one of \(N\) sets of process parameters in \(\bar{D}\).
% Added: (where \(h\) is normalized to one for both training and inference)

Training data was generated based on members of \(\bar{D}\). The MRFT response of each system within \(\bar{D}\) was simulated multiple times according to the process diagram shown in Fig. \ref{fig:sys_id_overview}. Measurement noise power \(\aleph\) was randomly varied between different simulations to add a regularization effect and prevent over-fitting \cite{Noh2017}. To further prompt robustness and generalization against varied testing conditions, simulations were carried out with varied values of forced input bias \(u_{0}\), which introduces asymmetry to the MRFT controller output. The values of \(u_{0}\) were limited to half the relay amplitude \(h\) of the MRFT controller as a reasonable bias magnitude in practical settings. Thirty simulations were performed for each candidate system in \(\bar{D}\) to produce a training set of size 6240. Five additional simulations per system were carried out to generate 1040 samples to be used as a verification set.

\begin{figure*}
\includegraphics[width=\textwidth]{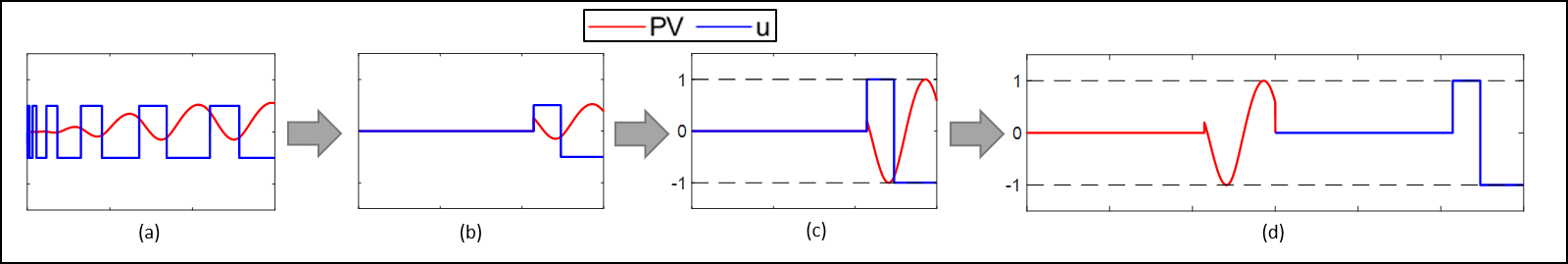} 
\caption{Pre-processing the DL classifier input vector: (a) The system's MRFT response is obtained and sampling time is adjusted to be 1ms. (b) A single cycle of the steady state oscillation is selected, zero-padding is applied elsewhere. (c) The response is zero-center and scaled to an amplitude of 1. (d) \(PV\) and \(u\) are concatenated to form a 1D vector }
\label{fig:dnn_preprocess}
\end{figure*}

The pre-processing steps undertaken to prepare the DL input data can be summarized as: sampling adjustment, cropping, zero-padding,  amplitude normalization, and concatenation. Fig. \ref{fig:dnn_preprocess} illustrates these pre-processing steps. Sampling period was fixed to be 1ms. The size of the input vector was set to be \(X \in  \mathbb{R}^{2\times 2260}\) to accommodate the response of the slowest system in \(\bar{D}\) (i.e. a period of 2.26s).

\begin{figure}
\includegraphics[width=\linewidth]{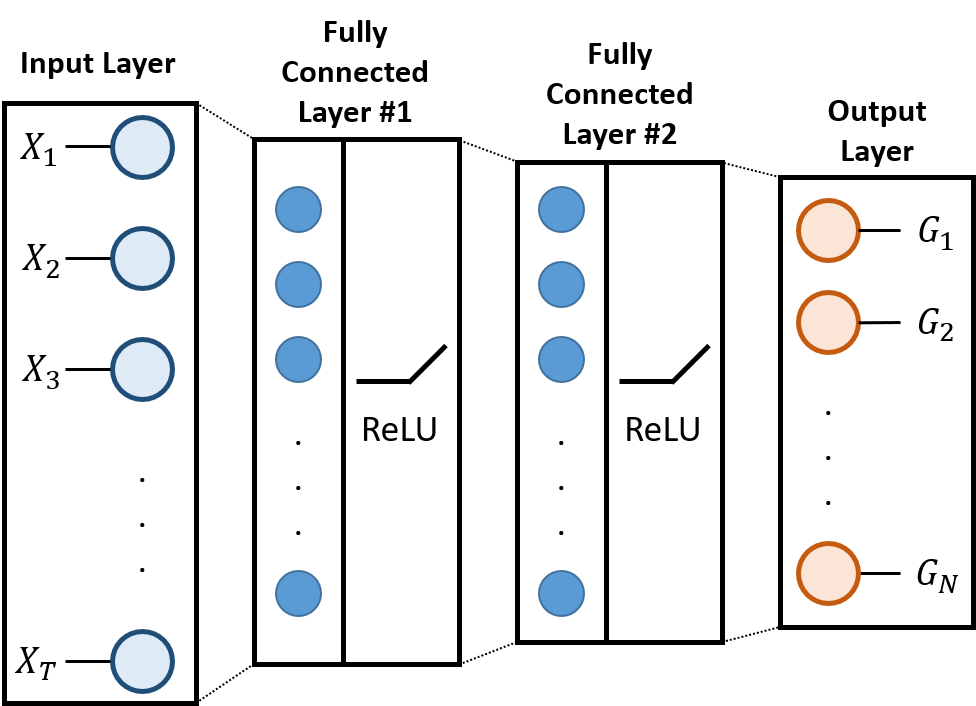} 
\caption{DL model architecture. The input vector length is 4520. Layer 1 contains 3000 neurons and Layer 2 contains 1000 neurons.}
\label{fig:DNN_structure}
\end{figure}

Fig. \ref{fig:DNN_structure} shows the structure of the developed deep learning model. The DL network consists of two hidden layers of size 3000 and 1000 respectively. This structure was chosen upon testing with several DL models of different depth and width up to four layers and 10000 neurons, as it showed the best performance with suitable performance on a signle-core processor. Convolutional  Neural Networks were also tested with no noticeable performance improvements. Rectified Linear Units (ReLU) were utilized as the activation function for both hidden layers due to its simplicity, reliability, and to avoid gradient vanishing \cite{Li2017}. Dropout is used after each layer for its regularization effect to avoid overfitting and prompt a noise rejection behavior \cite{Srivastava2014}. Batch normalization is applied to the outputs of the hidden layers to accelerate training and to add a slight regularization effect \cite{Ioffe2015}. The output layer consists of 208 neurons, one for each system in \(\bar{D}\).

The Softmax activation function and the Cross-entropy loss function are among the most utilized combinations for training deep learning networks. For the case of system identification, a conventional application of this combination lacks in the sense that the cost of incorrect classifications is identical regardless of the corresponding error in parameter space. To undermine this shortcoming, we introduce a modified formulation of the Softmax function. The modified formulation utilizes the joint cost function presented in \eqref{eq_performance_deterioration} to add a meaningful discrepancy to the cost of misclassification. For the \(i^{th}\) logit \(a_{i}\) in the output layer corresponding to a process \(G_{i}\), the modified Softmax probability \(p_{i}\) is introduced as:

\begin{equation} \label{eq_modified_softmax}
p_{i} = \frac{e^{\gamma_{iT}\cdot a_{i}}}{ \sum_{j=1}^{N}  e^{\gamma_{jT}\cdot a_{j}}}
\end{equation} 
where \(T\) is the class corresponding to the ground truth system \(G_{T}\), \(N\) is the size of the output layer, and \(\gamma_{iT}=1+J_{iT}\).\\

The DL network was trained using a Stochastic Gradient Descent approach with the cross-entropy loss function \(L=-\sum_{i=1}^{N} y_i \log{(p_i)}\), where \(y\) is a one-hot encoded vector that indicates the ground truth class \(T\). The partial derivative of \(L\) with respect to output layer logits \(a_{i}\) when using the Softmax function in \eqref{eq_modified_softmax} is calculated as:
\begin{equation} \label{eq_modified_softmax_backprop}
\frac{\partial L}{\partial a_{i}} = J_{iT} \times (p_{i} - y_{i})
\end{equation} 

For an exact derivation of \eqref{eq_modified_softmax_backprop}, readers can refer to Appendix \ref{app_modified_softmax_derive}.\\

Table \ref{tab:DNN_performance} demonstrates the classifier's performance on the verification set utilizing both the modified and conventional Softmax formulations. Although the DL network was used as a classifier, classification accuracy is not a suitable measure of the performance as it does not reflect the error between the predicted system \(G_{p}\) and \(G_{T}\). As such, controller performance deterioration \(J_{pT}\) is considered as a better and impartial evaluation criteria, especially as it was the basis for discretizing the system identification problem. 

\renewcommand{\arraystretch}{1.5}
\begin{table}[]
    \caption{DL verification set results on simulation data}
    \centering
    \begin{tabular}{|c|c|c|}
    \hline
            & \textbf{Modified Softmax} &  \textbf{Conventional Softmax} \\
        \hline
        Classification Accuracy & 38.46\% & 30.38\% \\
        \hline
        Average \(J_{pT}\) & 0.30\% & 0.41\% \\
        \hline
        Maximum \(J_{pT}\) & 5.03\% & 13.29\% \\
        \hline
        Minimum \(\phi_{m}\) & 13.73 & 4.89\\
        \hline
    \end{tabular}
    \label{tab:DNN_performance}
\end{table}

Results show that although classification accuracy is relatively low, the average joint cost \(J_{pT}\) is well below \(J^{*}=10\%\). The modified Softmax formulation results in a generally lower \(J_{pT}\) than the conventional Softmax function, particularly when comparing the maximum cost of misclassification. These results highlight a promising performance of the developed DL scheme in system identification applications under varied measurement noise \(\aleph\) and input bias \(u_{0}\). A single inference run of the DL model on a single core of an i5-6300U processor requires 5 ms, which reflects the suitability of the developed deep learning framework for real-time identification applications.

\section{results} \label{sec:results}

This section evaluates the effectiveness of the comprehensive system identification approach proposed in this paper. The testing approach follows the framework indicated by Fig. \ref{fig:sys_id_overview}; where the MRFT response of a dynamic system is obtained and passed to a DL model to predict the parameters of the PUT. Evaluation was performed using simulation data in addition to experimental tests on a UAV. Simulation results are compared against PEM as a well-established system identification method and the non-parametric tuning rules as an optimal controller design criteria.

\subsection{Simulation Results} \label{subsec:simulation_results}

Fifty different system parameter combinations were randomly sampled from the parameter space \(D\) to form a testing set \(\bar{D}_{test}\). Unlike the DL verification set in \ref{subsec:DNN_model}, these systems are not members of the discretized parameter space \(\bar{D}\), and are thus better suited to evaluate the generalization performance of the inclusive system identification solution. Testing data is generated by simulating the MRFT response for each system in \(\bar{D}_{test}\) under varied \(u_{0}\) and \(\aleph\). Data is then pre-processed and passed to the DL classifier to predict the parameters of a system \(G_{p}\).

The controller performance deterioration \(J_{pT}\) is utilized to evaluate the accuracy of identification. The mean and maximum deterioration values for the testing set are shown in Table \ref{tab:DNN_testing} using both the conventional Softmax function and the modified formulation in \eqref{eq_modified_softmax}. Both the average and maximum \(J_{pT}\) are within \(J^{*}=10\%\), which validates the proposed means of system identification. The modified Softmax formulation outperforms the conventional one as it results in lower controller performance deterioration and provides a larger margin for \(\phi_m\). These results demonstrate the generalization capabilities of the system identification DL framework to accommodate for the full parameter space \(D\) under different conditions of system bias and measurement noise.  

\renewcommand{\arraystretch}{1.5}
\begin{table}[]
    \caption{DL testing set \(D_{test}\) results on simulation data}
    \centering
    \begin{tabular}{|p{2.5cm}|c|c|}
        \hline
            & \textbf{Modified Softmax} &  \textbf{Conventional Softmax} \\
        \hline
        Average \(J_{pT}\) & 0.53\% & 0.75\% \\
        \hline
        Maximum \(J_{pT}\) & 3.51\% & 6.38\% \\
        \hline
        Minimum \(\phi_{m}\) & 15.53 & 12.15 \\
        \hline
    \end{tabular}
    \label{tab:DNN_testing}
\end{table}

\subsection{Comparison with the prediction error method}

\begin{figure}
\centering
\subfloat[]{\label{fig:sf1}
\includegraphics[width=7.5cm]{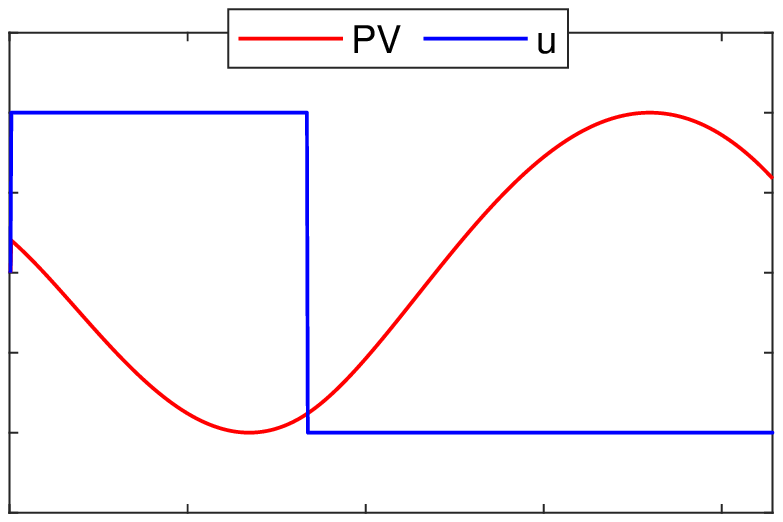}} \qquad

\subfloat[]{\label{fig:sf1}
\includegraphics[width=7.5cm]{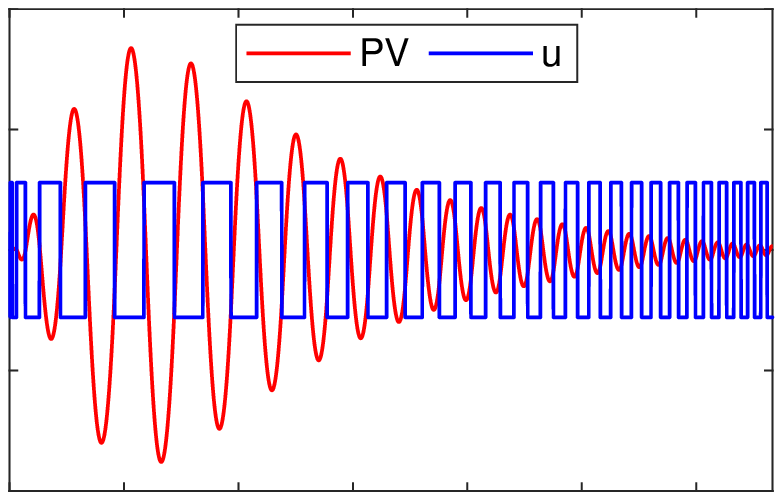}} \qquad
\caption{Sample of PEM input/output estimation data (a) Estimation Set-I with a single cycle of MRFT response (b) Estimation Set-II with multiple cycles of MRFT response with varied \(\beta\).}
\label{fig:PEM_estimationdata}
\end{figure}

The system identification performance of the proposed method was benchmarked against PEM using the same testing set \(\bar{D}_{test}\). PEM was implemented using Matlab's System Identification Toolbox \cite{matlabsysid} with Sequential Quadratic Programming as the search algorithm. To generate input/output estimation data for PEM, the closed-loop response of each system in \(\bar{D}_{test}\) was simulated under varied \(u_{0}\) and \(\aleph\). MRFT was chosen as the closed-loop controller is due to its guarantee of stability for all systems in \(D\). Accordingly, two different sets of estimation data were simulated for each PUT in order to assess PEM's performance against the amount of observation data. The first set (Estimation Set-I) consists of a single cycle of the MRFT response with the distinguishing phase \(\beta=-0.73\) of section \ref{findphase}. Fig. \ref{fig:PEM_estimationdata}-a shows a sample PEM estimation data of the first set, which is on par with the requirements of the DL approach proposed in this paper. The second set (Estimation Set-II) consists of 20 seconds of the simulated MRFT response with the \(\beta\) parameter continuously swept from \(\beta_{max}=-0.1\) to \(\beta_{min}=-0.9\); resulting in oscillations of different magnitude and frequency as shown in Fig. \ref{fig:PEM_estimationdata}-b. Finally, the most sensitive system in \(D\) for parameters variation which is \(G_{initial}=[0.1, 0.015, 0.2]\) was selected as the initial guess for PEM predictions in this study unless explicitly stated otherwise.

\renewcommand{\arraystretch}{2}
\begin{table}[h!]
  \begin{threeparttable}
    \caption{Comparison of system identification performance on simulation data for testing set \(\bar{D}_{test}\)}
    \centering
    \begin{tabular}{|p{2.3cm}|c|c|c|}
        \hline
            & 
            \multicolumn{2}{c|}{\textbf{PEM}} 
            &  \multirow{2}{*}{\textbf{DL}} \\ \cline{2-3}
            &\textbf{Estimation Set-I} & \textbf{Estimation Set-II} & \\
        \hline
        Average \(J_{pT}\) & \(10.82\%^{*}\)  &  \(4.41\%^{*}\) & 0.52\% \\
        \hline
        Maximum \(J_{pT}\) & unstable & unstable & 3.51\%\\
        \hline
        Number of unstable predictions & 5  & 1 & 0\\
        \hline
        Computation time per inference (s) & 1.8936  & 4.8740 & 0.005\\
        \hline
    \end{tabular}
    \begin{tablenotes}
     \small
     \item \vspace{0.1cm}\hspace{0.25cm} *: Excluding unstable predictions.
    \end{tablenotes}
    \label{tab:DNN_vs_PEM}
  \end{threeparttable}
\end{table}

Table \ref{tab:DNN_vs_PEM} compares PEM identification results against our DL approach in terms of controller performance deterioration \(J_{pT}\) and computation time per inference. Unstable predictions are defined as those that results in \(J_{pT}\) which grows to infinity with time (note that a condition to handle steady-state error is used). PEM results on Estimation Set-I show that unlike the deep learning approach, PEM fails to generate reliable predictions of system parameters from just a single cycle of the MRFT response. Increasing the amount of observation data enhances the prediction accuracy of PEM as indicated by results on Estimation Set-II; but it requires significantly larger processing time and still does not satisfy the maximum deterioration target of \(J^{*}=10\%\).

By examining the cases where PEM fails to produce accurate estimations, two main factors affecting PEM performance were identified. The first of which is PEM's requirement of a good initial guess. Table \ref{tab:PEM_initial} shows PEM predictions for a system \(G_{T}:\{T_{prop}=0.02, T_{body}=0.3, \tau = 0.001\}\) with different initial guesses. PEM predictions differ significantly with the initial guess and do not consistently converge to a suitable solution. By contrast, our proposed approach does not require an initial guess or prior knowledge of the PUT to generate appropriate system identification results.

\renewcommand{\arraystretch}{2}
\begin{table}[]
    \caption{PEM system identification results with different initial guesses for a simulated process \(G_{T}:\{T_{prop}=0.02, T_{body}=0.3, \tau = 0.001\}\)}
    \centering
    \begin{tabular}{|c|c|c|}
        \hline
            \multirow{2}{*}{\begin{tabular}{@{}c@{}}\textbf{Initial Guess} \\ $G_{p}$: \{$T_{prop}$, $T_{body}$, $\tau$\}\end{tabular}} &          \multicolumn{2}{|c|}{\textbf{Controller Performance Deterioration $J_{pT}$}}  \\ \cline{2-3}
            & \textbf{Estimation Set-I} & \textbf{Estimation Set-II} \\
        \hline 
        \(\{0.015, 0.2, 5\times10^{-4}\}\) & 0.0\% &  6.7\% \\
        \hline
        \(\{0.015, 0.2, 0.1\}\) & unstable & 21.2\% \\
        \hline
        \(\{0.3, 2, 0.0005\}\) & unstable & unstable \\
        \hline
        \(\{0.3, 2, 0.1\}\) & unstable & 26.0\% \\
        \hline
        \(\{0.15, 1, 0.05\}\) & unstable & 23.9\% \\
        \hline
    \end{tabular}
    \label{tab:PEM_initial}
\end{table}

The second factor disturbing PEM predictions is the input bias \(u_{0}\). Table \ref{tab:PEM_bias} compares identification results for a system \(G_{T}:\{T_{prop}=0.02, T_{body}=0.3, \tau = 0.001\}\) under different values of input bias using PEM and the suggested deep learning method. Referring to the results in Table \ref{tab:PEM_bias}, PEM predictions worsen as input bias get larger; by contrast, no significant differences are observed using our deep learning system identification method. 

\renewcommand{\arraystretch}{2}
\begin{table}[]
    \caption{Comparison of system identification performance on simulation data for process \(G_{T}:\{T_{prop}=0.02, T_{body}=0.3, \tau = 0.001\}\) with varied input bias}
    \centering
    \begin{tabular}{|c|c|c|c|}
        \hline
            \multirow{3}{*}{\textbf{Input Bias \(u_0\)}} & \multicolumn{3}{c|}{\textbf{Controller Performance Deterioration \(J_{pT}\)}} \\ 
            \cline{2-4}
            & \multicolumn{2}{c|}{\textbf{PEM}} &  \multirow{2}{*}{\begin{tabular}{@{}c@{}}\textbf{DL} \\ \textbf{Identification}\end{tabular}} \\
            \cline{2-3}   
            & \textbf{Estimation Set-I} &
            \textbf{Estimation Set-II}  &
             \\
        \hline
        0.0 & 0.0\% & 0.0\% & 0.0\% \\
        \hline
        \(-0.1 \times h_{mrft}\) & 0.0\% &  0.0\% & 0.0\% \\
        \hline
        \(-0.2 \times h_{mrft}\) & 0.0\% &  11.13\% & 0.0\% \\
        \hline
        \(-0.3 \times h_{mrft}\) & 0.0\% & 15.95\% & 0.0\% \\
        \hline
        \(-0.4 \times h_{mrft}\) & 7.49\% & 16.09\% & 0.0\% \\
        \hline
    \end{tabular}
    \label{tab:PEM_bias}
\end{table}

\subsection{Comparison with non-parametric tuning rules}

In order to assess the significance of the deep learning system identification solution for controller synthesis, it was benchmarked against the non-parametric tuning rules of \cite{Chehadeh2019} as it has comparable data and computational time requirements. These tuning rules were used to infer optimal PD controllers for each system in \(\bar{D}_{test}\). Using these rules, an average performance deterioration of 1.67\% and a maximum of 13.29\% were observed on the testing set. By comparison, the approach presented in this paper reduces these deterioration values to a third as indicated in Table \ref{tab:DNN_testing}. These performance improvements would be difficult to observe in practice as both approaches result in a low performance deterioration value.

The main advantage the work presented in this paper holds over the non-parametric tuning rules lies in its scalability to various controller structures and system models. The proposed approach identifies the dynamic model parameters, which enables the design of a wide set of controllers fitting to specific practical and performance requirements; as opposed to the PID structure limitation of the tuning rules. Accurate knowledge of the model parameters can further be utilized in designing other sub-systems such as: trajectory generation, state estimation, or multi-loop cascaded controllers. The presented solution can be extended to any parametric system identification or controller tuning problem with minimal modifications to the approach given the constraint on the number of unknown parameters. In contrast, adapting the tuning rules for different model structures would require extensive theoretical adjustments and analysis.  

\subsection{Experimental Results}

The approach presented in this paper was implemented to independently identify the altitude and attitude dynamics of a UAV. We utilized the Quanser QDrone as the testing platform for our experiments. The onboard IMU data was fused with Optitrack's motion capture system to estimate the pose of the drone. The procedure summarized in Fig. \ref{fig:sys_id_overview} is applied to a single control loop of the QDrone to identify its underlying process parameters and optimal PD controller. It must be noted that an optimal controller is initially designed offline for each system in \(\bar{D}\) to form a lookup table of optimal controller parameters. During operation, the process parameters are identified in real-time and optimal controller parameters are selected from the pre-designed lookup-table.

One advantage of our approach is the guarantee of stability during the identification phase by the MRFT controller \cite{Boiko2013}. Therefore, the identification procedure can be safely carried out without the need for prior knowledge of the underlying system dynamics. To illustrate this feature, identification of the altitude dynamics was done with the UAV starting from the ground with no PD controller. A generalized controller consisting of the summation of a MRFT controller and an integrator was used to elevate the UAV and excite stable oscillations around a predefined set-point. The objective of the integrator action is to counter the gravitational force. The following equation illustrates the quadrotor takeoff controller:
\begin{equation}\label{eq_integrator_takeoff}
    u_{i}(t) =  
    \\
    \left\{
    \begin{array}{l l}
    k_{i} \int (z_{ref} - z) dt \; &\text{if}\;z < z_{ref}\,\text{or}\, \dot{z} < \dot{z}_{max}
    \\
    u_{i}(t-) \quad &\text{otherwise}\; 
    \end{array}
    \right.
\end{equation}
%% Chehadeh: changed \frac{z_{ref}}{2}
where \(k_i\) is a constant gain, \(z\) and \(\dot{z}\) are the altitude and altitude change rate respectively, \(z_{ref}\) is the set point for altitude, \(\dot{z}_{max}\) is the maximum allowed altitude rate, and \(u_i(t-)\) is the previous controller output.

\begin{figure}
\includegraphics[width=\linewidth]{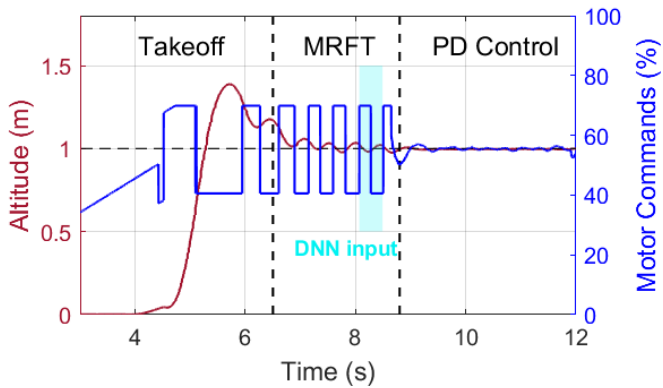} 
\caption{UAV experimental results for the altitude control loop. In the takeoff phase the algorithm presented in \eqref{eq_integrator_takeoff} was used. Note that MRFT takes a few oscillations to reach steady state. A single oscillation at steady state is selected as the input to the trained DL classifier.}
\label{fig:exp_height}
\end{figure}

Once a steady state oscillation is acquired, the identification scheme takes place using the pre-trained DL classifier. Based on the identified process parameters, an optimal PD controller is inferred and applied to the plant. Fig. \ref{fig:exp_height} shows the altitude and the controller action during the end-to-end identification and control process. The proposed take-off method was capable of stably lifting the UAV while simultaneously exciting oscillations. The DL network then identified the process parameters as \(G_{h}:\{T_{prop}=0.0321, T_{body}=1.6886, \tau = 0.0237\}\). Accordingly, the ISE optimal controller parameters were selected as \(C^{*}_{h}:\{K_{p}=59.0220, K_{d}=9.0356\}\). The controlled system response demonstrates a stable and smooth performance. In the absence of a ground truth system, this favorable controller performance indicates the effectiveness of the presented identification technique in targeting realistic control applications. 

The identification experiment for the altitude dynamics was repeated with a payload of 400g attached to the drone; which corresponds to a 30\% increase in the drone's mass. The optimal PD controller parameters identified under the increase in mass were \(\{K_{p}=69.9732, K_{d}=11.0002\}\), which shows a reasonable inflation over the nominal parameters in \(C^{*}_{h}\).

\begin{figure}[h]
\includegraphics[width=\linewidth]{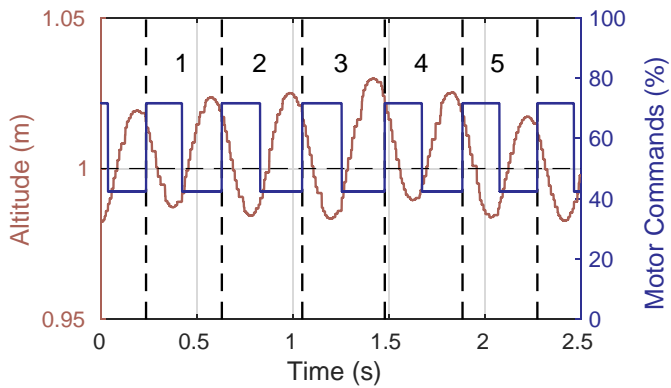} 
\caption{Multiple cycles of the experimental MRFT response for the UAV height control loop.}
\label{fig:exp_height_multiple_cycles}
\end{figure}

To assess the precision of the proposed identification scheme, the DL framework was tested with multiple cycles of the MRFT response as indicated in Fig. \ref{fig:exp_height_multiple_cycles}. Each cycle serves as an independent input to the DL model, which predicts system parameters accordingly. Table \ref{tab:DNN_confusion_matrix} shows the cross-performance deterioration matrix among the multiple identified systems from multiple steady-state oscillations corresponding to a single MRFT run. The maximum joint cost observed was \(1.97\%\) despite noticeable noise and variations among subsequent cycles of the response, which illustrates the precision and noise rejection capability of the trained DL classifier.  

\renewcommand{\arraystretch}{2}
\begin{table}[h]
    \caption{cross performance deterioration matrix showing controller performance deterioration \(J_{ij}\) for DL system identification of different cycles of the experimental MRFT response in Fig. \ref{fig:exp_height_multiple_cycles}}
    \centering
    \begin{tabular}{|c||c|c|c|c|c|}
        \hline
            \backslashbox{cycle i}{cycle j} & 1 & 2 & 3 & 4 & 5 \\
        \hline
        \hline
        1 & - & 0.19\% & 0.0\% & 0.9\% & 0.0\% \\
        \hline
        2 & 0.77\% & - & 0.0\% & 1.58\% & 0.01\% \\
        \hline
        3 & 1.25\% & 0.27\% & - & \textbf{1.97\%} & 0.35\% \\
        \hline
        4 & 0.0\% & 1.03\% & 0.45\% & - & 0.0\% \\
        \hline
        5 & 0.14\% & 0.64\% & 0.23\% & 0.95\% & - \\
        \hline
    \end{tabular}
    \label{tab:DNN_confusion_matrix}
\end{table}

\begin{figure*}
\centering

\includegraphics[width=\linewidth]{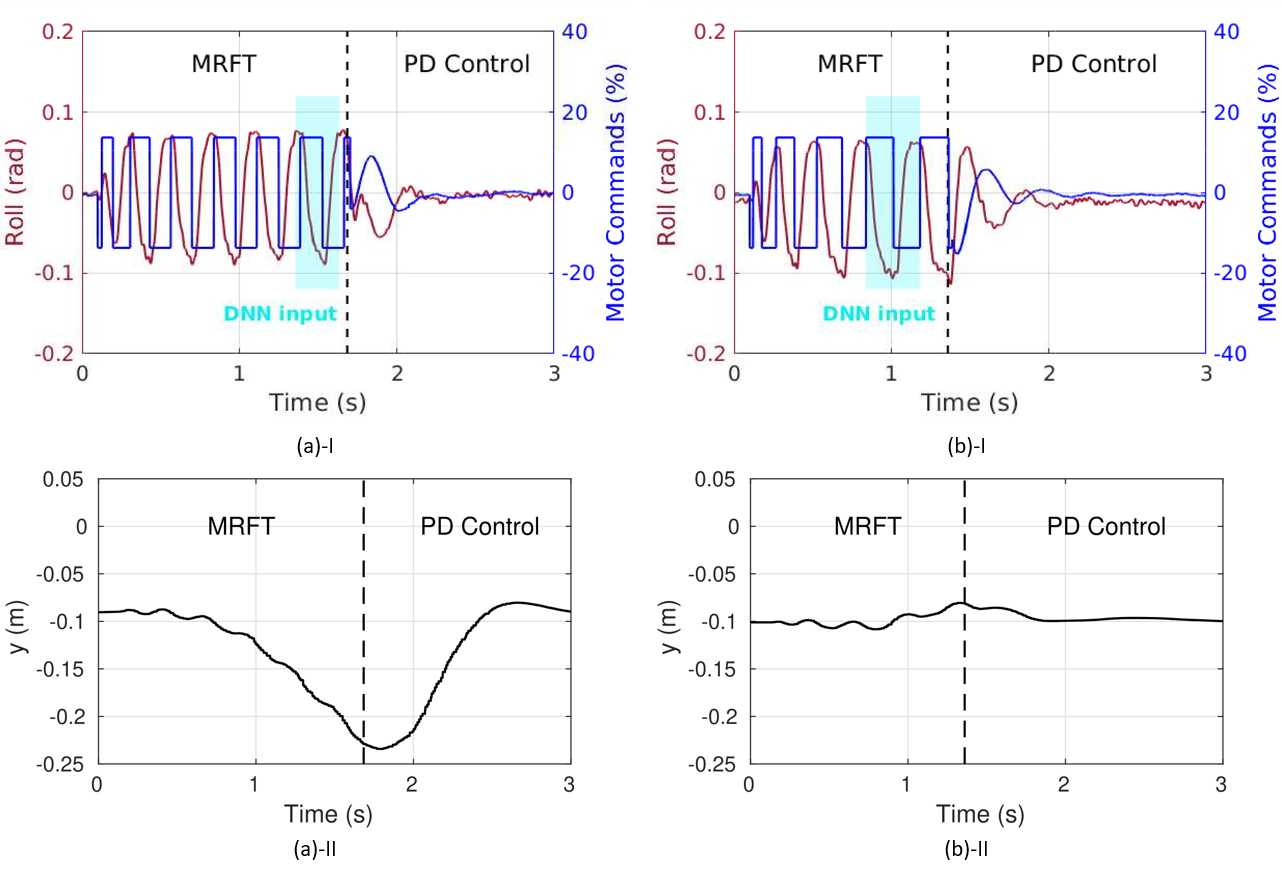}

\caption{UAV experimental results for the roll control loop: (a) Without bias compensation for the outer-loop, inherent system biases cause a drift in the lateral position of the drone. (b) With bias compensation, lateral drift is minimized. Proper system and controller parameters for the roll dynamics were identified in both scenarios.}
\label{fig:exp_roll_nobias}
\end{figure*}

The identification process was further applied to the roll dynamics of the QDrone. The closed loop system starts with a sub-optimal stabilizing controller \(C(s)\) that enables that UAV to safely take-off; the system identification procedure is then applied and an optimal controller \(C^{*}(s)\) is inferred accordingly. It must be noted that the outer-loop position controller is disabled during the identification phase to maintain a constant reference value for the MRFT controller. As a result, inherent system biases can cause a lateral drift in the drone's position. To undermine the translational drift, a bias compensation technique was implemented where the output of the roll channel's sub-optimal controller \(C(s)\) is filtered and stored prior to initiating the identification procedure. The filtered output is then used to offset the MRFT controller during the system identification process. 

Fig. \ref{fig:exp_roll_nobias} shows the results of the roll channel identification experiments during the subsequent identification and control phases with and without compensating for biases. When applying the bias compensation technique, the DL classifier identified the process parameters as \(G_{r}:\{T_{prop}=0.02, T_{body}=1.6889, \tau = 0.0121\}\), and the corresponding optimal PD controller parameters were \(C^{*}_{r}:\{K_{p}=1.1000, K_{d}=0.0985\}\). The PD control stage shows a positive response of the attitude dynamics to the designed controller and no substantial drift in the drone's position was observed. For the experiment where bias is not compensated for, the identified optimal controller parameters were \(C^{*}_{r^{-}}=\{K_{p}=1.1082, K_{d}=0.1013\}\). \(C^{*}_{r^{-}}\) is almost identical to \(C^{*}_{r}\), and would cause a practically negligible performance deterioration of less than \(0.2\%\) when applied to \(G_{r}\). This demonstrates the robustness of the developed identification framework under different testing conditions, which can widen its scope of application to a broad range of practical control problems. 

\subsection{supplementary material}

For better visualization of the experimental results, readers are encouraged to refer to the supplemental video in \cite{ayyad_2020}, which better highlights the performance of the proposed identification and control framework when applied to altitude and attitude loops of a quadrotor UAV. In addition to showing the capability of devising high-performance controllers in real-time, the video demonstrates the robustness of these controllers  to several artificial disturbances applied during operation time. These disturbances mimic practical conditions a UAV might encounter during a mission flight; and include weight changes, external nudges and induced wind speeds up to 5 m/s. The UAV sustains stability and performance despite the extreme conditions; which highlights the robust capabilities of the presented methodology and its applicability to practical identification and control problems.

\section{Conclusion} 
\label{sec:conclusion}

This paper introduced a novel approach for linear systems identification of a degree up to SOIPTD. The proposed method combines MRFT and DL to obtain a distinctive frequency response from an unknown plant, and map this response to a set of process parameters. The design of the identification procedure is presented as finding the distinguishing phase for a family of processes with the same model structure. System identification is then approached as a classification problem by using the principle of controller performance deterioration. Subsequently, a DL classifier is constructed and trained on noisy simulated MRFT responses. The end-to-end identification process takes place online requiring few seconds of observation data and microsecond level inference. 

The suggested approach was verified through simulation and experimentation. Experiments were carried out to identify the altitude and attitude dynamics of a UAV. Results show the effectiveness of the presented techniques by demonstrating stability in the adaptation phase, accuracy of identification, and real-time computation capabilities. The proposed method was bench-marked against PEM and the optimal tuning rules; and exhibited advantages in accuracy, robustness to input biases, less observation data requirements, and faster inference. All of which makes the presented approach applicable to a wide set of practical control problems.

For future work, we aim to evaluate the proposed identification scheme for higher order systems by extending the techniques of finding the distinguishing phase and parameter space discretization to higher dimensions.

\appendices
\section{Derivative of the modified softmax function with the cross-entropy loss function}
\label{app_modified_softmax_derive}

The derivative of the Softmax probability \(p_{i}\) from \eqref{eq_modified_softmax} with respect to the \(k^{th}\) logit can be obtained using the quotient rule:

\begin{multline}
    \label{eq_softmax_guotient_rule}
    \frac{ \partial p_{i} }{ \partial a_{k} } = \\
    \frac{ \frac{\partial}{\partial a_{k}} (e^{\gamma_{iT}\cdot a_{i}})\cdot\sum_{j=1}^{N}  e^{\gamma_{jT} \cdot a_{j}} 
    - 
    \frac{\partial}{\partial a_{k}}(\sum_{j=1}^{N}  e^{\gamma_{jT} \cdot a_{j}})\cdot e^{\gamma_{iT}\cdot a_{i}}}
    {(\sum_{j=1}^{N}  e^{\gamma_{jT} \cdot a_{j}})^{2}} 
    \\
    = 
    \frac{ \frac{\partial}{\partial a_{k}} (e^{\gamma_{iT}\cdot a_{i}})\cdot\sum_{j=1}^{N}  e^{\gamma_{jT} \cdot a_{j}} 
    - 
    \gamma_{kT} e^{\gamma_{kT} \cdot a_{k}} \cdot e^{\gamma_{iT}\cdot a_{i}}}
    {(\sum_{j=1}^{N}  e^{\gamma_{jT} \cdot a_{j}})^{2}} 
\end{multline}

In the case \(i = k\):
\begin{multline}
    \frac{ \partial p_{i} }{ \partial a_{k} } 
    = \frac{ \partial p_{k} }{ \partial a_{k} }
    \\
    =
    \frac{ \gamma_{kT} e^{\gamma_{kT}\cdot a_{k}}\cdot\sum_{j=1}^{N}  e^{\gamma_{jT} \cdot a_{j}} 
    - 
    \gamma_{kT} e^{\gamma_{kT} \cdot a_{k}} \cdot e^{\gamma_{kT}\cdot a_{k}}}
    {(\sum_{j=1}^{N}  e^{\gamma_{jT} \cdot a_{j}})^{2}} 
    \\
    =
    \gamma_{kT} \frac
    {e^{\gamma_{kT}\cdot a_{k}}}
    {\sum_{j=1}^{N}  e^{\gamma_{jT} \cdot a_{j}}}
    \cdot
    \frac
    {\sum_{j=1}^{N}  e^{\gamma_{jT} \cdot a_{j}} - e^{\gamma_{kT}\cdot a_{k}}}
    {\sum_{j=1}^{N}  e^{\gamma_{jT} \cdot a_{j}}}
    \\
    =
    \gamma_{kT} p_{k} (1-p_{k})
\end{multline}

Alternatively, for \(i \neq k\), \(\frac{\partial}{\partial a_{k}} (e^{\gamma_{iT}\cdot a_{i}}) = 0 \). (\ref{eq_softmax_guotient_rule}) hence simplifies as:

\begin{multline}
    \frac{ \partial p_{i} }{ \partial a_{k} } = 
    \frac
    {0 - 
    \gamma_{kT} e^{\gamma_{kT} \cdot a_{k}} \cdot e^{\gamma_{iT}\cdot a_{i}}}
    {(\sum_{j=1}^{N}  e^{\gamma_{jT} \cdot a_{j}})^{2}} 
    \\ =
    \gamma_{kT}
    \frac
    {-e^{\gamma_{kT} \cdot a_{k}}}
    {\sum_{j=1}^{N}  e^{\gamma_{jT} \cdot a_{j}}}
    \cdot 
    \frac
    {e^{\gamma_{iT}\cdot a_{i}}}
    {\sum_{j=1}^{N}  e^{\gamma_{jT} \cdot a_{j}}}
    =
    - \gamma_{kT} \cdot p_{k} \cdot {p_{i}}
\end{multline}

Therefore, the derivative of the modified Softmax formulation is:

\begin{equation}
    \label{eq_modified_softmax_derivative}
    \frac{\partial p_{i}}{\partial a_{k}} = 
    \\
    \left\{
    \begin{array}{l l}
    -\gamma_{kT} \cdot p_{k} (1-p_{k})\quad &\text{if}\; i = k
    \\
    - \gamma_{kT} p_{k} {p_{i}} \quad &\text{if}\; i \neq k
    \end{array}
    \right.
\end{equation}

The cross-entropy function \(L = - \sum_{i=1}^{N} y_{i} \log(p_{i})\) has the following derivative with respect to the \(k^{th}\) logit:

\begin{equation}
    \label{eq_crossentropy_derivative}
    \frac{\partial L}{\partial a_{k}} =
    - \sum_{i=1}^{N} y_{i} \frac{1}{p_{i}} \frac{\partial p_{i}}{\partial a_{k}}
\end{equation}
Where \(y\) is  a  One-Hot  encoded  vector that  points  out  the
ground truth class \(T\). 

By plugging \(\frac{\partial p_{i}}{\partial a_{k}}\) from (\ref{eq_modified_softmax_derivative}) to (\ref{eq_crossentropy_derivative}), the backpropagation term \(\frac{\partial L}{\partial a_{k}}\) becomes:

\begin{multline}
    \frac{\partial L}{\partial a_{k}} =
    - y_{k} \gamma_{kT} (1 - p_{k}) 
    + \sum_{\substack{i\neq k}} y_{i} (\gamma_{kT} \cdot p_{k})
    \\
    =
    \gamma_{kT} [-y_{k} + y_{k} p_{k} + \sum_{\substack{i\neq k}} y_{i} p_{k}]
    \\
    =
     \gamma_{kT} [-y_{k} + p_{k} (y_{k} + \sum_{\substack{i\neq k}} y_{i})]
\end{multline}

As \(y\) is a one-hot encoded vector, \(y_{k} + \sum_{\substack{i\neq k}} y_{i} = 1\). The backpropogation term hence becomes:

\begin{equation}
    \frac{\partial L}{\partial a_{k}} = 
    \gamma_{kT} (y_{k} - p_{k}) 
\end{equation}

\section*{Acknowledgment}

We would like to thank Quanser for the generous support and technical assistance. We also thank Mohamad Wahbah and AbdulRahaman Al-Marzooqi for their technical assistance with the experiments. This publication is based upon work supported by the Khalifa University of Science and Technology under Award No. RC1-2018-KUCARS.

\bibliographystyle{IEEEtran}

\bibliography{main.bib}

\end{document}